\documentclass{PoS}

\usepackage{amsmath}

\title{Multiple interactions and generalized parton distributions}

\ShortTitle{Multiple interactions and generalized parton distributions}

\author{\speaker{Markus Diehl}\\
        Deutsches Elektronen-Synchroton DESY, 22603 Hamburg, Germany\\
        E-mail: \email{markus.diehl@desy.de}}

\abstract{Multiple parton interactions in a single proton-proton collision
  are expected to play an important role for many observables at LHC.  To
  a large part their phenomenological description relies on rather simple
  and physically intuitive assumptions.  We report on an investigation of
  multiple hard interactions in QCD, which aims at identifying to which
  extent this simple description arises from theory and where it needs to
  be extended.  An approximate connection with generalized parton
  distributions may help elucidate some aspects of multiple-interaction
  dynamics.}

\FullConference{XVIII International Workshop on Deep-Inelastic Scattering
  and Related Subjects, DIS 2010\\
  April 19-23, 2010\\
  Firenze, Italy}

\newcommand{\half}{\tfrac{1}{2}}
\newcommand{\tvec}[1]{\mathbf{#1}}
\newcommand{\ms}{\mskip 1.5mu}

\begin{document}


\section{Introduction}

In proton-proton collisions at very high energies one can have events in
which several partons in one proton scatter on partons in the other proton
and produce particles of large transverse momentum or large mass.  The
effects of such multiparton interactions average out in sufficiently
inclusive observables, which can be described by conventional
factorization formulae that involve a single hard scattering.  Multiple
interactions do however change the structure of the final state and may be
important for many analyses at LHC \cite{Jung:2009eq}.  A brief review on
the subject can be found in \cite{Sjostrand:2004pf}.

Phenomenological estimates of multiparton dynamics rely on models that are
physically intuitive but rather simplified, whereas a systematic
description in QCD remains a challenge.  The present contribution reports
on some steps towards this goal.  The discussion is limited to tree-level
graphs, which already exhibit important features.  Loop corrections, in
particular soft gluon exchange, add a further layer of complexity and are
essential for determining if and in which form factorization holds.  For
details and further discussion we refer to \cite{Diehl:2010}.


\section{Cross section formula and multiparton distributions}

An example process where multiparton interactions contribute is the
production of two electroweak gauge bosons ($W$, $Z$ or $\gamma^*$ with
large virtuality) in kinematics where the transverse momenta of the bosons
are small compared with their mass or virtuality.  Since we are interested
in the details of the final state, we keep the cross section differential
in the transverse boson momenta.  For the corresponding process with a
single boson, a powerful theory in terms of transverse-momentum dependent
parton densities has been developed \cite{Collins:1981uk}, which one may
hope to generalize to the case of multiple hard scattering.

\begin{figure}[b]
\includegraphics[height=0.31\textwidth]{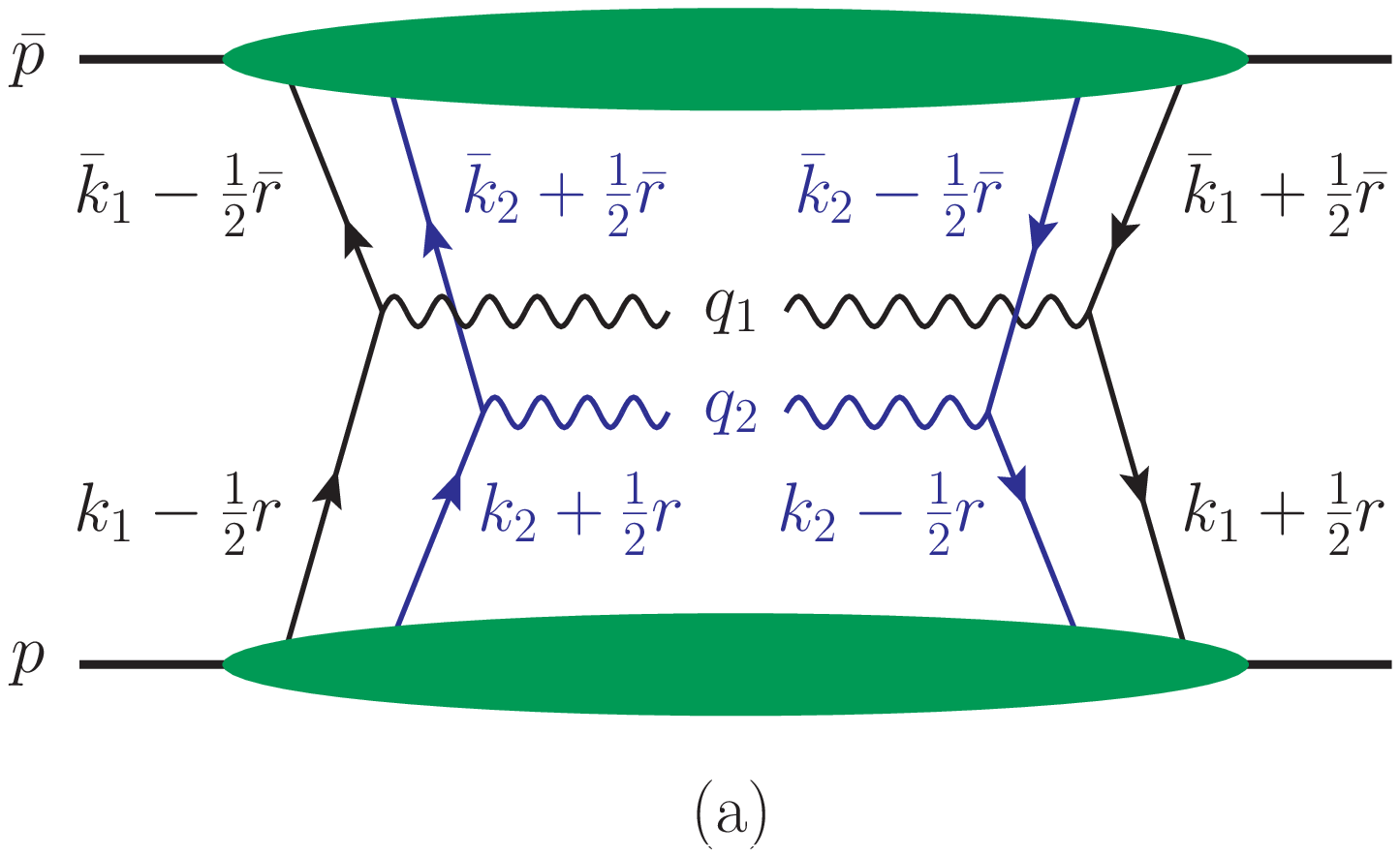}
\hfill
\includegraphics[height=0.31\textwidth]{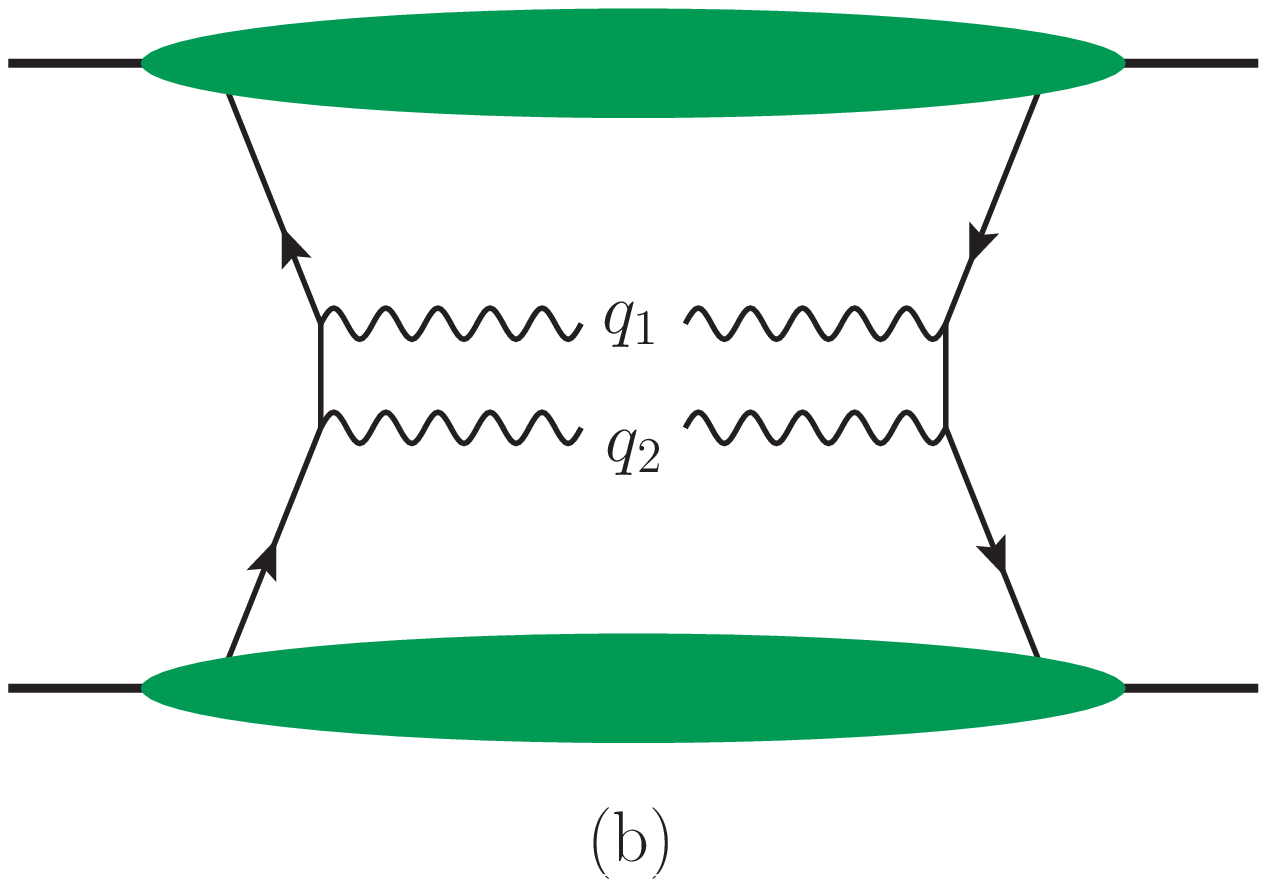}
\caption{\label{fig:scatter} Graphs for the production of two gauge bosons
  by double (a) or single (b) hard scattering.}
\end{figure}

Figure~\ref{fig:scatter}a shows a graph with two hard-scattering
subprocesses.  The transverse momentum of each produced boson is the sum
of the transverse momenta of a quark and an antiquark.  With two partons
emitted by each proton the transverse parton momenta are in general
\emph{not} equal in the scattering amplitude and its conjugate, as
illustrated in the figure.  Momentum conservation implies $\tvec{r} +
\bar{\tvec{r}} = \tvec{0}$ for the differences between the parton momenta
in the amplitude and its conjugate.  Performing a Fourier transform
w.r.t.\ $\tvec{r}$ and $\bar{\tvec{r}}$ one finds that their Fourier
conjugate variables are equal, $\tvec{y} = \bar{\tvec{y}}$.

Making the usual kinematic approximations in the hard-scattering
subprocesses (i.e.\ neglecting small momentum components compared with
large ones) one obtains
\begin{align}
  \label{X-section}
& \frac{d\sigma}{\prod_{i=1}^2 dx_i\, d\bar{x}_i\; d^2\tvec{q}_i}
  \;\bigg|_{\text{Fig.~\protect\ref{fig:scatter}}a}
= \frac{1}{S}\,
  \sum_{\genfrac{}{}{0pt}{1}{a_1,\ms a_2
         = \ms q,\ms \Delta q, \delta q}{%
       \bar{a}_1,\ms \bar{a}_2
         = \ms\bar{q},\ms \Delta\bar{q}, \delta\bar{q}}}
  \hat{\sigma}_{a_1 \bar{a}_1}(\hat{s} = x_1 \bar{x}_1 s)\;
  \hat{\sigma}_{a_2 \bar{a}_2}(\hat{s} = x_2\ms \bar{x}_2\ms s)
\nonumber \\[0.2em]
&\quad\times
  \biggl[\, \prod_{i=1}^{2} 
     \int d^2\tvec{k}_i\; d^2\bar{\tvec{k}}_i\;
     \delta^{(2)}(\tvec{q}_i - \tvec{k}_i - \bar{\tvec{k}}_i) \biggr]\,
  \int d^2\tvec{y}\;
  F_{a_1,\ms a_2}(x_i, \tvec{k}_i, \tvec{y})\, 
  F_{\bar{a}_1,\ms \bar{a}_2}(\bar{x}_i, \bar{\tvec{k}}_i, \tvec{y})
\end{align}
with $S=2$ if the produced bosons are identical and $S=1$ if they are not.
$\hat{\sigma}_{a_i \bar{a}_i}$ denotes the hard-scattering cross section
for single-boson production, and the kinematic variables are $s =
(p+\bar{p})^2$, $x_i = (q_i \bar{p}) /(p \bar{p})$ and $\bar{x}_i = (q_i
p) /(\bar{p} p)$.  The definition of double-parton distributions closely
resembles the one for the transverse-momentum dependent distribution of a
single quark \cite{Collins:1981uk}.  For the two-quark distribution in the
proton with momentum $p$ we have
\begin{align}
  \label{dist-def}
F_{a_1,\ms a_2}(x_i, \tvec{k}_i, \tvec{y})
 &= \biggl[\, \prod_{i=1}^2
       \int \frac{dz_i^-\, d^2\tvec{z}_i^{}}{(2\pi)^3}\;
       e^{i\ms x_i^{}\ms z_i^- p^+ - i\ms \tvec{z}_i^{} \tvec{k}_i^{}}
    \biggr]\, 2p^+ \!\!\! \int dy^-
\nonumber \\
 & \quad \times
    \big\langle p \big|\,
    \bar{q}\bigl( -\half z_2 \bigr)\, 
         \Gamma_{a_2}\, q\bigl( \half z_2 \bigr)\,\,
    \bar{q}\bigl( y - \half z_1 \bigr)\,
         \Gamma_{a_1}\, q\bigl( y + \half z_1 \bigr) \big| p \,\big\rangle
    \Big|_{z_1^+ = z_2^+ = y^+_{\phantom{1}} = 0} \;,
\end{align}
where we use light-cone coordinates $u^\pm = (u^0 \pm u^3) /\sqrt{2}$ for
each four-vector $u$.  One may regard $\tvec{k}_i$ as the ``average''
transverse momentum of each quark and $\tvec{y}$ as their ``average''
transverse distance from each other, where the ``average'' refers to the
scattering amplitude and its conjugate.  Regarding the transverse
coordinates $F(x_i, \tvec{k}_i, \tvec{y})$ has the structure of a Wigner
distribution \cite{Hillery:1984}, which depends on both momentum and
position variables for each particle.  (This does not contradict the
uncertainty principle since Wigner distributions are not probability
densities.)  A similar application of this concept has been discussed for
generalized parton distributions in \cite{Belitsky:2003nz}.

Integrating \eqref{X-section} over the transverse boson momenta
$\tvec{q}_1$ and $\tvec{q}_2$ one obtains
\begin{align}
  \label{X-section-coll}
\frac{d\sigma}{\prod_{i=1}^2 dx_i\, d\bar{x}_i}
  \;\bigg|_{\text{Fig.~\protect\ref{fig:scatter}}a}
&= \frac{1}{S}\, 
  \sum_{\genfrac{}{}{0pt}{1}{a_1,\ms a_2
         = \ms q,\ms \Delta q, \delta q}{%
       \bar{a}_1,\ms \bar{a}_2
         = \ms\bar{q},\ms \Delta\bar{q}, \delta\bar{q}}}
  \hat{\sigma}_{a_1 \bar{a}_1}(\hat{s} = x_1 \bar{x}_1 s)\;
  \hat{\sigma}_{a_2 \bar{a}_2}(\hat{s} = x_2\ms \bar{x}_2\ms s)
\nonumber \\[0.2em]
&\quad\times
  \int d^2\tvec{y}\;
  F_{a_1,\ms a_2}(x_i, \tvec{y})\, 
  F_{\bar{a}_1,\ms \bar{a}_2}(\bar{x}_i, \tvec{y}) \,.
\end{align}
The transverse-momentum integrated distribution $F_{a_1,\ms a_2}(x_i,
\tvec{y}) = \int d^2\tvec{k}_1\; d^2\tvec{k}_2\; F_{a_1,\ms a_2}(x_i,
\tvec{k}_i, \tvec{y})$ may be interpreted as the probability density for
finding two quarks with momentum fractions $x_1$ and $x_2$ at a relative
transverse distance $\tvec{y}$ in the proton.  The form
\eqref{X-section-coll} has long been known (see
e.g.~\cite{Paver:1984ux,Mekhfi:1983az}) and underlies most
phenomenological models for multiparton interactions.

For each quark there are three relevant Dirac matrices in
\eqref{dist-def},
\begin{align}
\Gamma_{q}          &= \half \gamma^+ \, , &
\Gamma_{\Delta q}   &= \half \gamma^+ \gamma_5 \, , &
\Gamma_{\delta q}^j &= \half i \sigma^{j +} \gamma_5
   ~~~\text{with $j=1,2$} \,,
\end{align}
which respectively project on unpolarized, longitudinally polarized and
transversely polarized quarks.  Note that polarized two-parton
distributions appear even in an unpolarized proton, since they describe
spin correlations between the two partons.  For small but comparable $x_1$
and $x_2$ one may well have sizeable spin correlations between the two
quarks (which are close in rapidity for $x_1 \sim x_2$), even if there is
little correlation between the polarizations of a quark and the proton
(which are far apart in rapidity).  The relevance of such correlations in
multiparton interactions was pointed out already in \cite{Mekhfi:1983az}
but has to our knowledge never been included in phenomenological
estimates.

Note that if parton spin correlations are sizeable they can have a strong
impact on observables.  For the production of two gauge bosons one can
easily see that the product $F_{\Delta q, \Delta q}\; F_{\Delta\bar{q},
  \Delta\bar{q}}$ of longitudinal spin correlations enters the cross
section with the same weight as the unpolarized term $F_{q,q}\,
F_{\bar{q},\bar{q}}$.  One also finds that product $F_{\delta q, \delta
  q}\; F_{\delta\bar{q}, \delta\bar{q}}$ of transverse spin correlations
give rise to a $\cos(2\varphi)$ modulation in the angle $\varphi$ between
the decay planes of the two bosons and thus affects the distribution of
final-state particles.

The formulae given so far have ignored the color structure of the
multiparton distributions.  The quark lines with momenta $k_1 \pm \half r$
in Fig.~\ref{fig:scatter}a can couple to a color singlet (as in
single-parton distributions) but they can also couple to a color octet,
provided that the lines with momenta \mbox{$k_2 \pm \half r$} are coupled to a
color-octet as well.  Such color-octet distributions contribute to the
multiple-scattering cross section, as was already pointed out in
\cite{Mekhfi:1983az}.  A more detailed discussion will be given in
\cite{Diehl:2010}.


\section{Power behavior}

It is easy to determine the power behavior of the cross section formula
\eqref{X-section} for two hard scatters.  The hard-scattering cross
sections $\hat{\sigma}$ behave like $1/Q^2$, where $Q^2 \sim q_1^2 \sim
q_2^2$ denotes the size of the large squared mass or virtuality of the
gauge bosons.  With transverse momenta $\tvec{q}_1 \sim \tvec{q}_2$ of
generic hadronic size $\Lambda$ we find
\begin{align}
  \label{power-double}
\frac{d\sigma}{\prod_{i=1}^2 dx_i\, d\bar{x}_i\; d^2\tvec{q}_i}
  \;\bigg|_{\text{Fig.~\protect\ref{fig:scatter}}a} 
& \sim\, \frac{1}{Q^4 \Lambda^2} \,,
\end{align}
where we have used that the two-parton distributions scale like $F \sim
1/\Lambda^2$ and that the typical transverse distance $\tvec{y}$ between
the partons is of order $1/\Lambda$.  The \emph{same} power behavior as in
\eqref{power-double} is obtained for the case where both bosons are
produced in a single hard scattering, as shown in Fig.~\ref{fig:scatter}b.
For the cross section differential in the transverse boson momenta,
multiple hard interactions are therefore \emph{not} power suppressed.

The situation changes when one integrates over $\tvec{q}_1$ and
$\tvec{q}_2$.  In the double-scattering mechanism both transverse momenta
are restricted to be of size $\Lambda$, but for a single hard scattering
one has $|\tvec{q}_1 + \tvec{q}_2| \sim \Lambda$ whereas the individual
transverse momenta can be as large as $Q$.  Because of this phase space
effect one has
\begin{align}
\frac{d\sigma}{\prod_{i=1}^2 dx_i\, d\bar{x}_i}
  \;\bigg|_{\text{Fig.~\protect\ref{fig:scatter}}a} 
& \sim\, \frac{\Lambda^2}{Q^4} \, , & 
\frac{d\sigma}{\prod_{i=1}^2 dx_i\, d\bar{x}_i}
  \;\bigg|_{\text{Fig.~\protect\ref{fig:scatter}}b} 
& \sim\, \frac{1}{Q^2} \,.
\end{align}
In the transverse-momentum integrated cross section, multiple hard
scattering is therefore only a power correction.  This is required for the
validity of the usual factorization formulae, which describe only the
single-scattering contribution.


\section{Connection with generalized parton distributions}

A simple ansatz for modeling two-parton distributions is to write them as
a product of single-parton distributions, thus neglecting correlations
between the two partons.  This provides a starting point for
phenomenology, even though one may not expect such an approximation to be
very precise.  A way to implement this ansatz for the distributions
$F(x_i, \tvec{k}_i, \tvec{y})$ is to insert a sum $\sum_{X} |X\rangle\,
\langle X|$ over all physical states between the two bilinear operators in
\eqref{dist-def}.  If one \emph{assumes} that the proton state is dominant
in this sum, one has
\begin{align}
\big\langle p \big|\,
    \bigl( \bar{q}_{\,x_2}\, q_{\,x_2} \bigr)\,
    \bigl( \bar{q}_{\,x_1}\, q_{\,x_1} \bigr)
    \big| p \big\rangle
& \,\approx\,
\int \frac{dp'^+}{2 p'^+}\, \frac{d^2\tvec{p}'}{(2\pi)^3}\;
\big\langle p \big|\,  \bar{q}_{\,x_2}\, q_{\,x_2}
    \big| p' \big\rangle \;
\big\langle p' \big|\, \bar{q}_{\,x_1}\, q_{\,x_1}
    \big| p \big\rangle  \,,
\end{align}
where the subscripts $x_1$, $x_2$ indicate the momentum fraction
associated with each field and spinor indices have been omitted for
brevity.  After the Fourier transform in \eqref{dist-def}, momentum
conservation imposes $p'^+ = p^+$ but one still has $\tvec{p}' \neq
\tvec{p}$.  The two-parton distribution $F(x_i, \tvec{k}_i, \tvec{y})$ is
thus approximated by a product of two \emph{generalized} parton
distributions with zero skewness $\xi$.  For transverse-momentum
integrated distributions one obtains a simple representation $F(x_i,
\tvec{y}) = \int d^2\tvec{b}\, f(x_2, \tvec{b})\, f(x_1,
\tvec{b}+\tvec{y})$ in terms of impact parameter dependent parton
densities \cite{Burkardt:2002hr}.

The same method can be applied to the color-octet distributions mentioned
above.  Rearranging the quark operators as
\begin{align}
\half
\bigl( \bar{q}_{\,x_2}\ms \lambda^a\, q_{\,x_2} \bigr)\,
\bigl( \bar{q}_{\,x_1}\ms \lambda^a\, q_{\,x_1} \bigr)
&= {}- \bigl( \bar{q}_{\,x_1}\, q_{\,x_2} \bigr)\,
     \bigl( \bar{q}_{\,x_2}\, q_{\,x_1} \bigr)
 - \tfrac{1}{3}
     \bigl( \bar{q}_{\,x_2}\, q_{\,x_2} \bigr)\,
     \bigl( \bar{q}_{\,x_1}\, q_{\,x_1} \bigr)  \, ,
\end{align}
where $\lambda^a$ are the Gell-Mann matrices, one can insert proton states
between the operators in parentheses on the r.h.s.  For the first term
this leads to generalized parton distributions with \emph{nonzero}
skewness $\xi$, since the fields coupled to a color singlet carry
different longitudinal momenta.

Although it is only approximate, the connection with generalized parton
distributions that can be measured in exclusive reactions --- pointed out
already in \cite{Frankfurt:2003td} --- will hopefully help to better
understand at least some features of multiple hard interactions.


\end{document}